\documentclass[floatfix,twocolumn,showpacs,preprintnumbers,amsmath,amssymb,aps,prb]{revtex4-1}

\usepackage{color} \usepackage{graphicx}
\usepackage{dcolumn}
\usepackage{bm}
\usepackage{ulem}

\newcommand{\pr}{Pr$^{3+}$} 
 \newcommand{\thf}{$^3$H$_{4}$}
 \newcommand{\odt}{$^1$D$_{2}$}
 
 \newcommand{\er}{Er$^{3+}$}

 \newcommand{\YSO}{Y$_2$SiO$_5$}

\newcommand{\LAWO}{La$_2$(WO$_4$)$_3$}

\begin{document}

\title{Hyperfine characterization and coherence lifetime
    extension in Pr$^{3+}$:La$_{2}$(WO$_4$)$_3$}

\author{Marko \surname{Lovri\'c}}
\email[]{marko.lovric@tu-dortmund.de} 
\author{Philipp Glasenapp}
\author{Dieter Suter} 
\affiliation{Technische Universit\"at Dortmund,
  Fachbereich Physik, 44221 Dortmund, Germany} 
  
\author{Biagio Tumino}
\author{Alban Ferrier} 
\author{Philippe Goldner}
\email[]{philippe-goldner@chimie-paristech.fr} 
\affiliation{Chimie ParisTech ENSCP, Laboratoire de Chimie de la Mati\'ere Condens\'ee
  de Paris(LCMCP), 75005 Paris, France\\ UPMC Univ Paris 06, 75005 Paris, France\\
  CNRS, UMR 7574, 75005 Paris, France}

\author{Mahmood Sabooni}
\author{Lars Rippe}
\author{Stefan Kr\"oll}
\affiliation{Department of Physics, Lund Institute of Technology, P.O. Box 118, S-221 00 Lund, Sweden}
\date{\today}

\begin{abstract}
Rare-earth ions in dielectric crystals are interesting candidates for storing quantum states of photons.
A limiting factor on the optical density and thus the conversion efficiency is the distortion introduced in the crystal
by doping elements of one type into a crystal matrix of another type.
Here, we investigate the system Pr$^{3+}$:La$_{2}$(WO$_4$)$_3$, where the similarity of the ionic
radii of Pr and La minimizes distortions due to doping.
We characterize the praseodymium hyperfine interaction of the ground state
($^3$H$_4$) and one excited state ($^1$D$_2$) and determine the spin Hamiltonian parameters 
by numerical analysis of Raman-heterodyne spectra, 
which were collected for a range of static external magnetic field strengths and orientations. 
On the basis of a crystal field analysis, we discuss the physical origin of the experimentally 
determined quadrupole and Zeeman tensor characteristics.
We show the potential for quantum memory applications by measuring the spin coherence lifetime in a magnetic field
that is chosen such that additional magnetic fields do not shift the transition frequency in first order.
Experimental results demonstrate a spin coherence lifetime of 158 ms - almost three orders of magnitude longer
than in zero field.
\end{abstract}

\pacs{42.50.Md, 76.30.Kg, 76.70.Hb, 76.60.-k, 03.67.Pp}



\maketitle

\section{\label{introduction}Introduction}
Rare-earth ion-doped crystals (REIC) have recently appeared as
promising solid state materials for quantum information processing.
In the field of quantum computing, achieved milestones include controlled phase gates
\cite{Longdell:2004p28} and single qubit arbitrary
rotation\cite{Rippe:2008p731}.
While these experimental results were performed on single-qubit and two-qubit systems,
scalable schemes have also been poposed \cite{Wesenberg:2007p217}. 
In the field of quantum memories, devices able to
faithfully store and release photonic quantum states have been proposed and implemented.
Using several different storage-recall protocols 
\cite{Nilsson:2005p7,Afzelius:2009p1246,Hetet:2008p728,Lauro:2009p1247},
high efficiency \cite{Hedges:2010p1245},
multiple photon storage with large bandwidth
\cite{Usmani:2010p1222,Bonarota2011} and entanglement
storage\cite{Clausen2011,Saglamyurek2011} were demonstrated in REICs.
These results rely on the shielding of the 4f electrons of the rare-earth ions
by closed shells, which reduces dephasing by the environment,
yielding long coherence lifetimes ($T_2$) at
liquid helium temperatures. For example, optical coherence lifetimes
of 4.4 ms have been observed in \er:\YSO{} [\onlinecite{Bottger:2009p1248}]. 

Even longer lifetimes have
been reported for rare-earth ion hyperfine transitions and accordingly, the
qubit in REIC based quantum computing and memories is generally
defined by selecting two ground state hyperfine levels. Optical
transitions are used to selectively address qubits or to transfer
coherences from the optical to the radio-frequency (RF) domain and
vice versa.  Coherence lifetimes can be extended
to 30 s for a ground state
hyperfine transition of \pr:\YSO{} at liquid helium
temperature \cite{Fraval:2005p122}.  This was achieved in two steps:
first, an external magnetic field was applied to the sample in order
to decouple one hyperfine transition from magnetic field fluctuations
due to host spin flips. As these are the main source of dephasing for
\pr{} hyperfine transitions, the zero field coherence lifetime of 500
$\mu$s was extended in this way to 82
ms\cite{Fraval:2004p115} and later to 860 ms
\cite{Fraval:2005p122}. The decoupling was achieved by
minimizing the transition energy dependence with respect to the magnetic
field. This condition is referred as ZEFOZ (Zero First Order Zeeman
shift) transitions.\cite{Longdell:2006p9} The coherence lifetime was
then further increased by RF decoupling pulses, using a modified version \cite{Fraval:2005p122} 
of the Carr-Purcell sequence originating from nuclear magnetic resonance.

\YSO{} is the most thoroughly
studied host in REIC quantum information processing.
It combines long coherence life
times, favored by its low magnetic moment density, mainly due to 
Y nuclear spins,  and high oscillator strengths. 
A disadvantage of Y-based host material is that doping with \pr{} or Eu$^{3+}$,
leads to relatively large inhomogeneous linewidths at high doping
concentrations, which limits the maximal achievable optical depth. 
This is an important concern in high efficiency quantum memories. \cite{Tittel:2010p1251} 
To overcome this limitation for
\pr, we proposed a La-based crystal, \LAWO. \pr{}
substitutes La$^{3+}$ in this material, both having very similar ionic radii 
($r_{La^{3+}}\,=\,1.18$ \AA, $r_{Pr^{3+}}=1.14$ \AA{}) \cite{Shannon:1969p729}. 
Compared to \YSO{} ($r_{Y^{3+}}= 1.02$ \AA{})\cite{Shannon:1969p729},
doping stress is reduced
and the inhomogeneous linewidth is 15 times smaller in this compound at high
\pr concentrations. 
However, the magnetic moment of lanthanum  (2.78
$\mu_B$) is much higher than that of Y$^{3+}$ (-0.14 $\mu_B$) and the
\LAWO{} magnetic moment density is 7.5 times higher than in \YSO. It
seems that this should be seriously detrimental to coherence lifetimes
but we measured a hyperfine lifetime of 250 $\mu$s by Raman-echoes
\cite{GuillotNoel:2009p1163}, which is only smaller by a factor of $\approx 2$
compared to the value in \YSO. This allowed us to measure narrow and
efficient electromagnetically induced transprency in this
material. \cite{Goldner:2009p726} This result also suggested that
REICs that could be useful for quantum information processing are not
limited to the few crystals with  very low magnetic moment
density. However, since applications require $T_2$ values in the ms range,
the techniques described above for increasing the coherence lifetime by
several orders of magnitude should be used. In this paper, we show
that using a ZEFOZ transition, hyperfine $T_2$ can reach \mbox{$158\pm7$ ms,}
corresponding to a 630-fold increase. It is therefore possible to strongly reduce the
influence of host spin flips even in the case of high magnetic moment
density.

ZEFOZ transitions appear at specific magnetic fields (magnitude and direction),
which can only be predicted if the system Hamiltonian and all of its parameters
are known with high precision.
In the present system, 
the $I=5/2$ nuclear spin of $^{141}$Pr (100 \% abundance) and the $C_1$
site symmetry result in a complicated hyperfine structure. 
We therefore used the approach of Ref.\ \onlinecite{Longdell:2002p29},
which consists in determining the spin Hamiltonian parameters by coherent
Raman-scattering before numerically identifying ZEFOZ transitions.  
Finally, the coherence lifetimes of the hyperfine transitions were 
measured by optically detected Raman-echoes.

\section{\label{theory}Model for the hyperfine interaction}
A good approximation for the Hamiltonian of many rare-earth doped
compounds is \cite{AAKaplyanskii:1987p1}
\begin{equation}
  \mathcal{H}_0=\left[ \mathcal{H}_{FI} + \mathcal{H}_{CF} \right] 
  + \left[ \mathcal{H}_{HF}+\mathcal{H}_Q+\mathcal{H}_Z+\mathcal{H}_z \right].
  \label{eq_H0}
\end{equation}
The first two terms, the free ion (including spin-orbit coupling) and
the crystal field Hamiltonians determine the energies of the
electronic degrees of freedom.  The terms in the second bracket,
consisting of the hyperfine coupling, the nuclear quadrupole coupling,
the electronic and the nuclear Zeeman Hamiltonian, lift the degeneracy
of the nuclear spin states.

The site symmetry ($C_1$) of our system is low enough that the
electronic states are nondegenerate.  As a result of this
``quenching'' of the electronic angular momentum, the electronic
Zeeman $\mathcal{H}_Z$ and hyperfine interaction $\mathcal{H}_{HF}$
contribute only as second order perturbations.  In this approximation,
the four last terms of Eq.\ (\ref{eq_H0}) can be well approximated by
a nuclear spin Hamiltonian \cite{Teplov:1968p739}:
\begin{eqnarray}
  \mathcal{H}_n 	&=&-g_{J}^2\mu_{B}^2 \vec{B} \cdot \mathbf{\Lambda} \cdot \vec{B} 
  \label{eq_H1} \\
  &  &	- \vec{B} \cdot \left( 2 A_J g_J \mu_B \mathbf{\Lambda} 
    + g_I \mu_N \mathbf{E} \right) \cdot \vec{I} \nonumber \\
  &  & + \vec{I} \cdot \left(\mathbf{P}- A_{J}^2 \mathbf{\Lambda} 
  \right) \cdot \vec{I}Ê\nonumber.
\end{eqnarray}
Here, $g_J$ is the Land\'{e} g-value, $\mu_B$ the Bohr magneton and
$\vec{B}$ the external magnetic field vector.  The $\mathbf{\Lambda}$
tensor
\begin{equation}
  \Lambda_{\alpha \beta} = 
  \sum_{n=1}^{2J+1} \frac{\langle 0 | J_\alpha | n \rangle \langle n | J_\beta | 0 \rangle}{E_n - E_0},
  \label{eq_lambda}
\end{equation}
is calculated by second order perturbation theory.  $\alpha$ and
$\beta$ denote the coordinate axes and the sum runs over the $2J+1$
states of the relevant crystal field multiplet.  The crystal field
levels of interest in the following are the lowest of the $^3$H$_4$
and $^1$D$_2$ multiplets and are written as $|0\rangle$.  Other
crystal fields levels of each of these multiplets are denoted by
$|n\rangle$ and $E_n$ represents the energy of that state.  $A_J$ is
the hyperfine constant of the $J$ multiplet, $g_I$ is the nuclear
$g$-factor, $\mu_N$ the nuclear magneton, $\mathbf{E}$ the 3x3 unity
matrix and $\vec{I}=\left(I_x, I_y, I_z\right)$ the vector of nuclear
spin operators.  $\mathbf{P}$ represents the pure nuclear quadrupole
tensor.  The operator $(-A_J^2\vec{I}\cdot\mathbf{ \Lambda
}\cdot\vec{I})$ has the same form as
  $\vec{I}\cdot\mathbf{P}\cdot\vec{I}$ and is therefore
conventionally referred to as the pseudoquadrupole interaction
\cite{Baker:1958p1161}.  We combine these two operators into an
effective quadrupole Hamiltonian $\vec{I}\cdot\mathbf{Q} \cdot
\vec{I}$.  In systems with low site symmetry, as in the present case,
the principal axis systems (PAS) of $\mathbf{P}$ and
$\mathbf{\Lambda}$ do not coincide in general.  The combined operator
$\mathbf{Q}$ therefore has again a different PAS.

Considering the first term of Eq.\ (\ref{eq_H1}), we note that it does
not depend on the nuclear spin.  It shifts all nuclear spin levels by
the same amount and we therefore neglect it in the following.
Furthermore, we introduce the abbreviation $\mathbf{M}= - \left(2 A_J
  g_J \mu_B \mathbf{\Lambda} + g_I \mu_N \mathbf{E} \right)$ for
the effective Zeeman coupling tensor.
We thus arrive at the following form of the nuclear spin Hamiltonian:
\begin{equation}
  \mathcal{H} = \vec{B}\cdot \mathbf{M}\cdot \vec{I} + \vec{I} \cdot\mathbf{Q}\cdot \vec{I}.
  \label{eq_H}
\end{equation} 
The tensors $\mathbf{M}$ and $\mathbf{Q}$ can be parametrized as
\cite{AAKaplyanskii:1987p1}
\begin{eqnarray}
  \mathbf{M} =	& R_M \cdot   \left[ 
    \begin{array}{ccc}
      g_{x}	& 		0 		& 0 \\
      0			& g_{y}		& 0 \\
      0			&				&g_{z}
    \end{array} 
  \right]  \cdot R_M^T \label{eq_M}  &\\
  \mathbf{Q} =	& R_Q \cdot  \left[ 
    \begin{array}{ccc}
      E-\frac{1}{3}D 	& 		0 		& 0 \\
      0			& -E-\frac{1}{3}D	& 0 \\
      0			&				& \frac{2}{3}D
    \end{array} 
  \right]  \cdot R_Q^T &,	
  \label{eq_Q}
\end{eqnarray}
where the $R_i=R(\alpha_i,\beta_i,\gamma_i)$ represent rotation
matrices and Euler angles, specifying the orientation between
$\mathbf{M}$ and $\mathbf{Q}$ tensors PAS ($(x',y',z')$ and
$(x'',y'',z'')$ respectively) relative to the laboratory-based
reference axis system $(x,y,z)$ (see section \ref{experiment}).  In
general, the $\mathbf{M}$ and $\mathbf{Q}$ principal axes are not
aligned and accordingly the $R_Q$ and $R_M$ matrices are not
identical.  The nuclear spin Hamiltonian of a given crystal field
level therefore depends on 11 parameters: $g_x,
g_y,g_z,\alpha_{{M}},\beta_{{M}},\gamma_{{M}},D, E,
\alpha_{{Q}},\beta_{{Q}},\gamma_{{Q}}$.

\begin{figure}
  \includegraphics[width=0.85 \columnwidth]{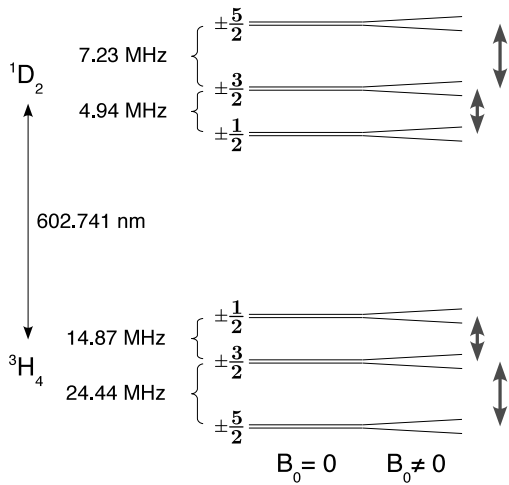}
  \caption{\label{fig_levels} Optical transition and hyperfine level
    structure of Pr$^{3+}$:La$_{2}$(WO$_4$)$_3$.  The order
    of the energy levels follows from previous work \cite{GuillotNoel:2009p1163}, whereas
    the transition frequencies could be measured with higher precision
    utilizing zero field Raman-heterodyne scattering within the
    present work.  The arrows on the right indicate 
    the range of hyperfine transitions excited by RF 
    in the separate experiments. }
\end{figure}

In zero magnetic field, the quadrupole interaction results in a
partial lifting of the nuclear spin states degeneracy.  
The corresponding structures for the \thf{} and \odt{} levels 
was determined by holeburning experiments
\cite{GuillotNoel:2007p2,GuillotNoel:2009p1163} and are shown in Fig.\
\ref{fig_levels}.  We label the levels by their projections onto the
$z''$ principal axes of the $\mathbf{Q}$ tensor, noting that
these states are not eigenstates of the nuclear spin Hamiltonian.  
Since the PAS of the $\mathbf{Q}$ tensors of different crystal field levels do not
coincide, their quantization axes are also different.  For the small
magnetic fields used to determine spin Hamiltonian parameters, the
hyperfine structure remains close to the zero field one, which allows
us to identify resonance lines with transitions between zero-field
states.

\section{\label{experiment}Experiment}
We used Raman-heterodyne scattering (RHS) to obtain hyperfine spectra
of both, the electronic ground state and the electronically excited
state.  RHS is a magneto-optic double resonance technique, that
requires a laser and an RF field \cite{Mlynek:1983p247,
  Wong:1983p1154}.  The laser light has multiple functions in the
scheme.  The first is to prepare a population difference between the
hyperfine levels by optical pumping or transfer to auxiliary states.
If the RF field is resonant with the hyperfine transitions, it creates
coherences between those states, which are transferred to coherences
in optical transitions by the light.  These optical coherences
represent electronic dipoles, which act as a source of a new optical
field that is shifted by the RF frequency with respect to the incident
laser frequency - the Raman-field.  As this field is emitted in the
same optical mode as the incident light, it can be measured by optical
heterodyne detection, using the transmitted laser beam as the local
oscillator.

We used a sample of high optical quality, grown by the
Czochralski method, containing 0.2\% at. Pr$^{3+}$.
The 5x5x5 mm crystal was mounted in an optical cryostat and cooled 
to liquid helium temperatures.  \LAWO{} forms a monoclinic
crystal with a $C2/c$ space group, identical to that of \YSO.
The La$^{3+}$ ions occupy only one crystallographic site of $C_1$
symmetry.  In each unit cell (containing 4 formula units), this site
 appears at 8 positions which are related by inversion, translation and $C_2$
 symmetries.  The $C_2$ axes are identical to the $C_2$ crystal
 symmetry axis, also denoted by $b$ in the following. 
 The $C_2$ symmetry divides the La positions into two groups of 4 ions,
 which behave differently, unless the magnetic
 field is perpendicular or parallel to the $b$ axis. These two groups
 are called sub-sites in the following. 
The crystal surfaces where polished perpendicular to the $(X,Y,Z)$ principal axes
of the optical indicatrix.  Optical back reflection at the $Z$-surface
and mechanical alignment of the X and Y surfaces was used to align the
crystal along our reference $(x,y,z)$ axes, defined by the static
magnetic field coils (see below).  Apart from alignment errors the
$(X,Y,Z)$ axes should be a replica of $(x,y,z)$, the laser propagating
along $Z$ and the $b$ axis expected to be closely aligned to the
laboratory frame $y$ axis.

A Coherent 899-21 dye laser, further stabilized by home-built
electronics with respect to intensity and frequency (linewidth $<$ 20 kHz),
served as light source.  It was tuned to a wavelength of 602.741 nm
(vac.), corresponding to the center of the transition involving the
lowest energy crystal field levels of the $^3$H$_4$ and $^1$D$_2$
multiplets (see Fig.\ \ref{fig_levels}).  Typical powers for the
scattering/heterodyne light were 0.3 to 2 mW focused, from a
collimated beam of 1.5 mm diameter, with a 300 mm lens into the
sample.  We generated the optical pulses and frequency chirps by
double pass acousto-optic modulator setups.

The RF fields were applied to the sample by a 10 turn 6 mm diameter
coil.  For continuous wave experiments (ground state), one side of the
coil was terminated by a 50 $\Omega$ load, the other was attached to
an RF driver running at a power level of 1 W.

\begin{figure}
  \includegraphics[width= \columnwidth]{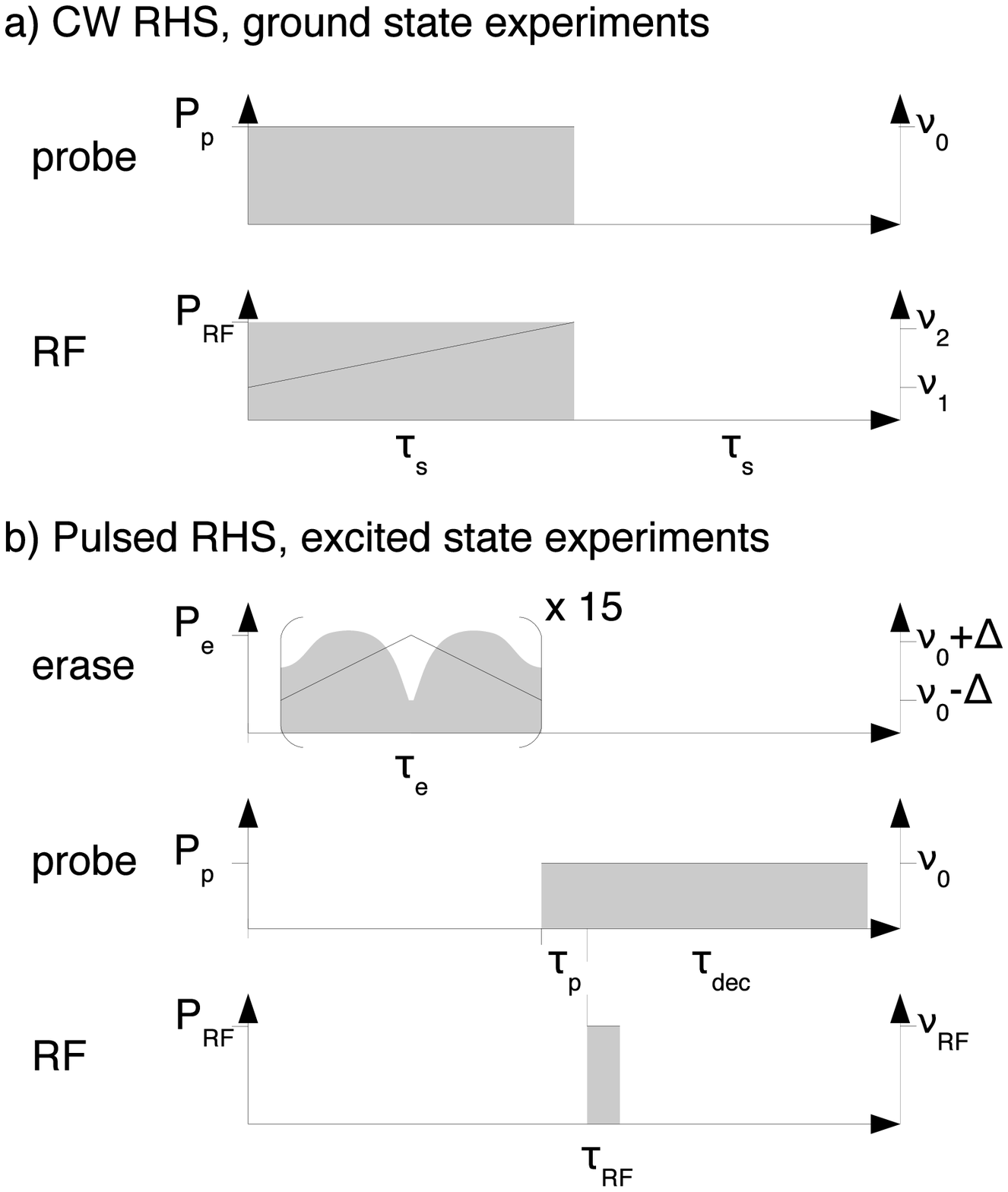}
  \caption{\label{fig_ppg} Pulse sequences. Black lines indicate frequencies and 
  grey areas the applied optical/RF powers. \\
a) CW (continuous wave) sequence for the ground state measurements.  
    A low power laser probe ($P_p$ = 0.3 mW) and a scanning frequency 
    RF ($P_{RF}\approx$ 1 W, $\nu_1$ = 7.4 and \mbox{$\nu_2$ = 22.4 MHz} for 
    $\left|\pm \frac{1}{2}\right> \leftrightarrow \left| \pm \frac{3}{2}\right>$ 
    or respectively 17.1 and \mbox{32.1 MHz} for $\left|\pm \frac{3}{2}\right> \leftrightarrow \left|\pm
    \frac{5}{2}\right>$) were applied
    to the sample to detect the spectra on the frequency encoded
    time-scale \mbox{($\tau_s$ = 50 ms).}
    The optimum temperature of the cold finger for this scheme was 4.5 K, 
    since still lower temperatures gave such slow hyperfine level relaxation 
    rates that it would be necessary to repump the hyperfine level population.\\
 b) Pulsed RHS-sequence for the excited state.  Probe and erase
    beam were overlapped in the sample at angle of 0.6$^\circ$.  To
    allow for higher repetition rate the chirped erase laser (\mbox{$\Delta$ = 64 MHz,} 
    $\tau_e$ = 10 ms, $P_e$ = 8-30 mW due to the frequency dependence of the AOM) 
    redistributed the populations.  Initial hyperfine
    population was created by the probe beam ($P_p$ = 1.2 mW,
    $\tau_p$ = 100 $\mu$s) and converted to coherences by an RF pulse
    ($\nu_{RF}$ = 4.94/7.23 MHz ($\left|\pm \frac{1}{2}\right>
    \leftrightarrow \left| \pm \frac{3}{2}\right>$ resp. $\left|\pm
      \frac{3}{2}\right> \leftrightarrow \left|\pm
      \frac{5}{2}\right>$), $P_{RF}$ = 214/287 W, $\tau_{RF}$ = 4 $\mu$s).
      For optical heterodyne detection the same probe 
      beam was left active for an additional time $\tau_{dec}$.
    The temperature of the cold finger for this scheme was 2.4 K. }
\end{figure}

For the exited state spectra, we used a pulsed RHS scheme.  Here the
coil was part of an appropriate tuned tank-circuit and typical RF
powers were 250 W, resulting in maximum signal amplitudes for RF-pulse
durations of {3-4 $\mu$s}.  Figure \ref{fig_ppg} shows the sequences
used for the two types of experiments.

The frequencies for the RF excitation and the shifting of the laser
frequency were generated by 48 bit / 300 MHz direct digital
synthesizers.  We controlled the timing of the pulses and
frequency-chirps by a wordgenerator with a resolution of 4 ns.
Detection of the heterodyne beat signal was accomplished by a 100 MHz
balanced photo receiver (Femto HCA-S), a phase sensitive
quadrature-detection demodulation scheme, appropriate analog and
digital filters and a digital oscilloscope.
 
The static magnetic field was created by a set of three orthogonal
Helmholtz coil pairs.  They are mounted outside the cryostat and their
coil-diameters range from 20 to 40 cm, providing a homogeneous field
over the sample volume in their center.  
With currents of about 10 A, each coil pair generates a
static magnetic field of about 8 mT. 
To control the field vector a computer control was set
up for the current sources of the Helmholtz coils.  To compensate
non-linearities and drifts, we used a set of three orthogonal Hall
probes as sensors for a computer-based feedback loop.  The absolute
error of the field components is $< 0.06$ mT and the relative linear
error for the static magnetic field is $< 0.3 \%$.  To minimize the
effect of small background fields (e.g. earths magnetic field) a small
compensation field   was used, which minimized the observed zero-field RHS line
splitting and also led to almost perfect destructive interference
\cite{Mitsunaga:1984p741,Mitsunaga:1985p211}.  
We used this compensation field as our zero-field reference in all measurements.

\begin{figure*}
  \includegraphics[width=2
  \columnwidth]{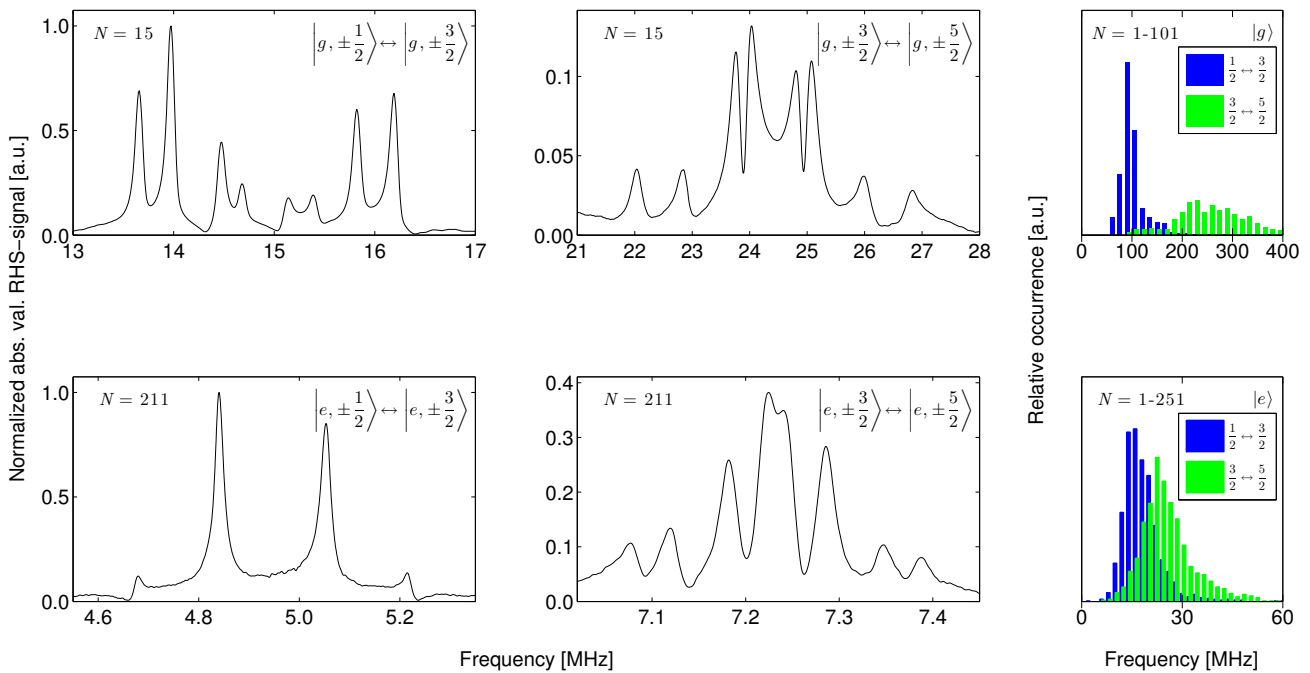}
  \caption{\label{fig_egspectra} (Color online) Representative ground state CW RHS
    and exited state pulsed RHS spectra.  Both spectra from the ground
    state ($\left| g,i \leftrightarrow j \right>$) are recorded at
    $N=15$, corresponding to $\vec{B}=( 2.60, -5.27,
      -5.76)$ mT.  For the excited state spectra, we used
    $N=211$, corresponding to $\vec{B}=( 4.62, 1.19, 4.42)$
    mT.  The normalization is relative to the largest line from ground
    or excited state spectra, respectively.  The ground state spectra
    shown here resolve all 8+8 possible RHS transitions, while in the
    excited state spectra not all lines are resolved.  The histograms
    on the right show the distributions of fitted full width at half
    maximum (FWHM) RHS-linewidths for all recorded data (see Section
    \ref{sourcedata}), plotted separately for the two hyperfine transitions ($\left| g/e,i \leftrightarrow j \right>$, 
    $i/j=\pm\frac{1}{2}/\pm\frac{3}{2}$ or $\pm\frac{3}{2}/\pm\frac{5}{2}$).
    For both, the ground and the excited state, 
    the $\pm\frac{3}{2} \leftrightarrow \pm\frac{5}{2}$ linewidths are bigger (see Section \ref{discussion}). 
    The mean linewidths are $\left| g, \pm\frac{1}{2} \leftrightarrow \pm\frac{3}{2}\right>\approx$ 105 kHz, 
    $\left| g, \pm\frac{3}{2} \leftrightarrow \pm\frac{5}{2}\right>\approx$ 301 kHz, 
    $\left| g, \mbox{all}\right>\approx$ 196 kHz and
    $\left| e, \pm\frac{1}{2} \leftrightarrow \pm\frac{3}{2}\right>\approx$ 18.3 kHz, 
    $\left| e, \pm\frac{3}{2} \leftrightarrow \pm\frac{5}{2}\right>\approx$ 26.3 kHz, 
    $\left| e, \mbox{all}\right>\approx$ 22.3 kHz.
    }
\end{figure*}

For an optimal determination of the Hamiltonian parameters, it is
important to sample different strengths and orientations of the
magnetic field.  In our experiments, we used a spiral on the surface
of an ellipsoid \cite{Longdell:2002p29}:
\begin{eqnarray}
  &
  \vec{B}(t) = \left( 
    \begin{array}{c}
      B_x \sqrt{1-t^2} \cdot \cos \left( 6 \pi t \right) \\
      B_y \sqrt{1-t^2} \cdot \sin \left( 6 \pi t \right) \\
      B_z \cdot t
    \end{array} 
  \right). \label{eq_B}
\end{eqnarray}
Here, we use
$$
t = -1 + \left( N-1 \right) \frac{2}{N_{tot}-1}, \quad
N=1,2,...,N_{tot}.
$$
to represent the discrete coordinate along the trajectory.  For the
ground state series, we measured $N_{tot}=$ 101 orientations, with
magnetic field amplitudes \mbox{$[B_x, B_y, B_z] = [7, 9, 8]$ mT} and for the
excited state we used $N_{tot}=$ 251 and $B_x=B_y=B_z=6.5$ mT.  Figure
\ref{fig_egspectra} shows some typical experimental spectra for the
ground- and excited state.

\section{\label{fitting}The fitting procedure}
In the case of Pr$^{3+}$:La$_{2}$(WO$_4$)$_3$, 11 spin Hamiltonian
parameters are necessary ($D$, $E$, $g_{x,y,z}$, $(\alpha, \beta,
\gamma)_{Q/M}$) to compute the line positions of excited or ground
state spectra for an arbitrary external magnetic field. These
calculations have to take into account the magnetically non-equivalent
\pr{} sites, which are related by the crystal $C_2/b$ axis,
corresponding to the $Y$-direction of the crystal's optical indicatrix
axis system $(X,Y,Z)$.  As this system may not be perfectly aligned to
the laboratory reference system $(x,y,z)$ (see section
\ref{experiment}), $(\alpha,\beta)_{C_2}$ angles give the $C_2$
orientation with respect to the $x$ and $z$ axes.  This gives a total
of 13 parameters that are to be extracted from the recorded series of
spectra.

\subsection{\label{sourcedata}Source data}
While it is possible to calculate directly the spectra from the
Hamiltonian, the resulting amplitudes and phases do not agree with the
experimentally observed values, since they depend strongly on
experimental details, such as cable lengths, tuning curves etc.  By
far the most precise parameters of the spectra are the positions of
the resonance lines.  We therefore extracted these in a first step and
used only the resonance frequencies as input to the actual fitting
procedure.

For the extraction of the line positions, we fitted the absolute-value
representation of the measured spectra to a series of gaussian-shaped
resonance lines.  To discriminate between real lines and noise or
artifacts, we considered additional parameters like peak height, line
width and noise level.  For most of the recorded data the procedure
worked satisfyingly, even for partially overlapping resonance lines.
The positions of lines that had very small amplitudes or could not be
fitted well due to strong overlapping were identified visually or
omitted.

\subsection{\label{sa}Fitting of Hamiltonian parameters}

After determining the resonance line positions, we fitted the
Hamiltonian parameters such that the resulting line positions agreed
with the experimental data.  Due to the large number of parameters and
their complicated interdependency, gradient-based algorithms are not
efficient, as they tend to stick to local minima.  In our case, the
combination of first running a probabilistic and then a direct search
yielded the best convergence to the global minimum.  As in the work of
Longdell \cite{Longdell:2002p29,Longdell:2006p9} we used simulated
annealing \cite{Kirkpatrick:1983p1155} for the first step.
When the preliminary results of this step appeared to converge towards
a global minimum, we switched to a pattern search algorithm \cite{Audet:2003p1270},
which polls meshes of adjacent points to find a better minimum.  This
reduced the necessity for an excessively long search at low
temperatures in the simulated annealing procedure and represents an
additional check for the quality of the minimum.

The implemented algorithm minimizes the root mean square (RMS)
deviation between the measured and the calculated line positions of
all spectra,
\begin{equation}
  f^{RMS}_k = \sqrt{\sum_{N=1}^{N_{tot}} \frac{\sum_{i=1}^{Q_N}\left ( \nu^{exp}_{iN}-\nu^k_{iN} \right )^2}{Q_N}} .
  \nonumber
\end{equation}
For each of the $N_{tot}$ spectra $Q_N$ line pairs, consisting of the
measured frequencies $\nu^{exp}_{iN}$ and the corresponding calculated
frequencies $\nu^k_{iN}$ were identified.  The index $i$ runs over all
lines in a single spectrum.  Accordingly all possible eigenvalue
differences ($\nu^k_{iN}$) for the given Hamiltonian, derived from
Eq.\ (\ref{eq_H}) using the given set of parameters of the $k^{th}$
iteration and the $N^{th}$ magnetic field orientation (Eq.\
(\ref{eq_B})), were calculated.  Not all possible lines are always
resolved or visible in the spectra (see e.g. Fig.\
\ref{fig_egspectra}).  Therefore, we matched theoretical to
experimental lines by minimizing the absolute deviations for this
spectrum; each experimental line was associated only with a single
theoretical line.  Lines that could not be assigned uniquely were not
considered in that step of the fitting procedure.  This mapping could
change during the fitting procedure, but in the final steps of the
search, the procedure resulted in an unambiguous assignment.

We initialized the simulated annealing algorithm with a random guess
for the parameters.  At each following iteration one of the
Hamiltonian parameters was randomly chosen and varied in an interval
determined by the current temperature.
If the variation resulted in a new minimum value of the RMS error
($f^{RMS}_{best}$), the new point was always accepted. 
If the variation was higher than the previous minimum, 
the algorithm calculated the Boltzman factor
\begin{equation}
  p_k = \exp \left( \frac{-\left(  f^{RMS}_k-f^{RMS}_{best}\right)}{k_{B_l} \cdot T_{k_l}}\right) \label{eq_Boltzmann}
\end{equation}
and compared it to a random number $p_r$ $\epsilon$ $\left[ 0,1Ê\right] $.  
The new point was accepted for $p_k > p_r$ and rejected otherwise.  
The RMS misfits $f^{RMS}_{k}$ and $f^{RMS}_{best}$ represent 
the deviation of the current iteration and
the best observed deviation of all previous iterations, respectively.
The temperatures $T_{k_l}$ are defined in the individual  $l$-th parameter units. 
Conversion factors $k_{B_l}$ ensured correct units and 
parameter independent scaling of the denominator in Eq.\ (\ref{eq_Boltzmann}).  
All temperatures were lowered continuously as the fitting improved. 
This was done for all $T_{k_l}$ in the same way, 
so that for simplicity we may speak in terms of a global temperature 
and omit the $l$-index where it is not necessary.   
At high temperatures the algorithm sampled a large parameter range.  Enabling it to
resettle in states of slightly higher misfits enables the algorithm to
trace a large search-space and to avoid getting stuck in a local
minima at the same time.  As the temperature was gradually lowered,
the parameter values became more confined in the vicinity of the
global minimum.

The whole procedure was implemented as a $\mbox{MATLAB}$ program, 
utilizing the Global Optimization Toolbox functions \textit{patternsearch} and
\textit{simulannealbnd}.

\subsection{\label{restrictions}Restrictions and Model}

We chose the initial temperature $T_{0}$ such that the corresponding
changes of the line positions were $\approx 10$ times the
inhomogeneous line widths.  Typically $k_{max} \approx 10^6$
iterations for the simulated annealing lead to reliable results.
Within the first 80\% of the iterations the temperature was gradually
lowered as $T_{k} \approx T_{0} \left ( 1-k/k_{max} \right ) ^2$.
Then $T_{k}$ was left constant at a value corresponding to a frequency
uncertainty of a fraction of the typical line-widths (a few kHz).

During the whole fit, the parameters were constrained.  As values for
$D$ and $E$ can be derived from hole-burning experiments
\cite{GuillotNoel:2007p2,GuillotNoel:2009p1163} their boundaries were
restricted to $\pm$10\% of the literature values.  For the
gyromagnetic ratios, the boundaries were chosen to $\pm$100\% of the
expected values of -(10-100) MHz/T\footnote{The negative sign of the gyromagnetic 
ratios preserves the consistence with the crystal field analysis in Sec.\ \ref{discussion}. 
It should be noted that this analysis assumes a higher crystal symmetry 
than the actual one of \pr:\LAWO{} and that RHS spectra are not sensitive to the sign 
of the gyromagnetic ratios, as explained later.} 
observed in similar systems.\cite{AAKaplyanskii:1987p1,Longdell:2002p29}
The Euler angles were allowed to vary over the whole definition range.  The probability for
choosing a specific parameter for the variation was proportional to
the (relative) size of its boundaries.  As indicated above, the new
value of this parameter was chosen by adding a random value out of the
interval $\pm T_{k_l}$.
If the new value did not lie within the
boundaries, the process was repeated.  The final direct search was set
up with the same boundaries and typically terminated after
$5\cdot10^4$ iterations.

The orthogonal axes of the B-field coils define the reference
coordinate system $(x,y,z)$.  Euler-angles and transformations refer
to this basis and are given in ``zyz''-convention
\cite{Goldstein:2002p1241} (see. Eq.\ (\ref{eq_EulerDef}) for
details).  As indicated before, we did not constrain the relative
orientation of the quadrupole $\mathbf{Q}$- and Zeeman
$\mathbf{M}$-tensor.  To avoid ambiguous results, we fitted all
measured spectra of a single electronic state simultaneously.  Due to
the two non-equivalent sites this results in a maximum of 16 lines per
spectrum.

Since the assignment of the resonance lines to the two sites is not
known, we had to fit both sites simultaneously.  Instead of fitting
$2\cdot11=22$ parameters, we used the fact that they are related by a
$C_2$ rotation.  Using Eq.\ (\ref{eq_H}) for site 1, we write the
Hamiltonian for site 2 as
\begin{eqnarray}
  \mathcal{H}_2  & = &  \vec{B} \cdot \left( R_{C_2} \mathbf{M}_1 R_{C_2}^T \right) \cdot \vec{I} 
  +  \vec{I} \cdot \left(R_{C_2} \mathbf{Q}_1 R_{C_2}^T \right) \cdot \vec{I}, 		\nonumber	 
\end{eqnarray}
with
\begin{eqnarray}
  & R_{C_2} =  R_C^T \cdot R_{\pi} \cdot R_C,& \nonumber \\
  & R_C  =  R \left( \alpha_{C_2}, \beta_{C_2}, 0 \right), \quad R_{\pi} = R \left( 180^\circ,0,0 \right).&
  \nonumber
\end{eqnarray}
The angles $\alpha_{C_2}$ and $\beta_{C_2}$ correspond to the spherical
coordinates of the $C_2$ axis in the laboratory system.

\section{\label{results}Results}

The underlying symmetry of the crystal field and the structure of the
spin Hamiltonian cause some ambiguity if only RHS spectra are used to
determine the Hamiltonian parameters.  The thesis of J. Longdell
\cite{JosephLongdell:2003p1216} provides a detailed review of relevant
symmetries that make it impossible to unambiguously determine all
Hamiltonian parameters from RHS spectra alone.  Important for our
investigation is the fact that the RHS spectra do not depend on the
signs of $D$, $E$ and the gyromagnetic factors $g_x$, $g_y$ and $g_z$.
In addition, different sets of Euler angles correspond to the same
tensor orientations.  As a consequence, different runs with random
initial values lead to apparently different solutions.  We checked
that these solutions are related by the symmetry operations mentioned
above and verified thus that we really found a unique global minimum.

\subsection{\label{g} Electronic ground state}

\begin{figure*}
  \includegraphics[width=2
  \columnwidth]{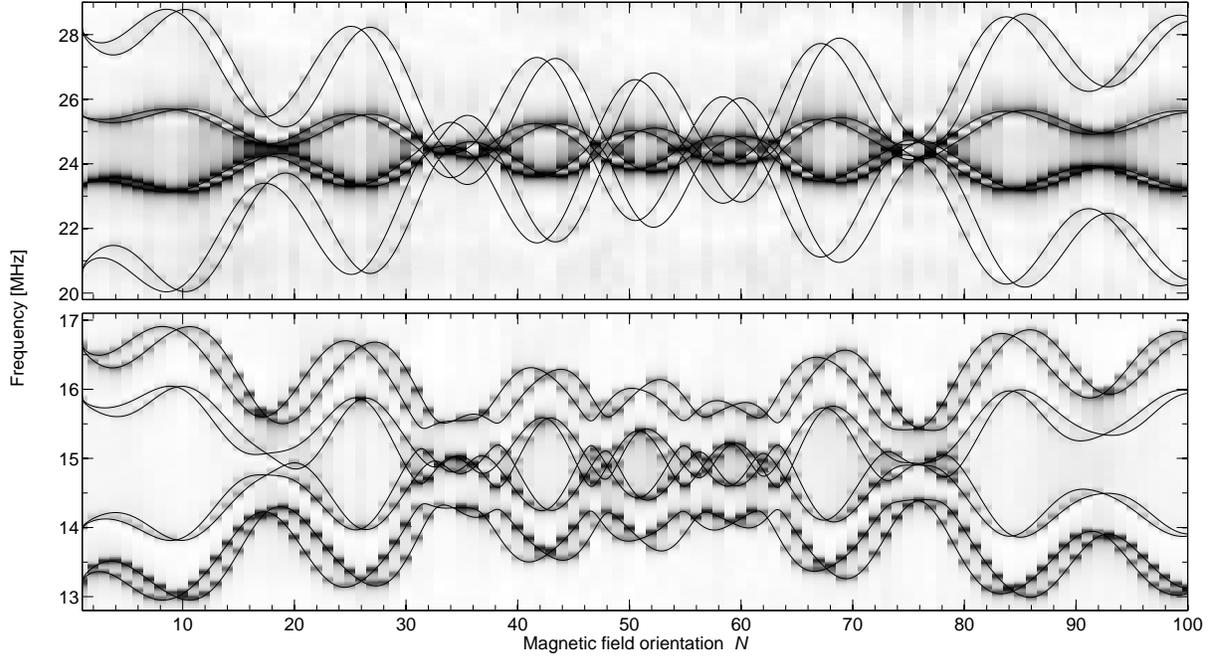}
  \caption{\label{fig_groundstate} Ground state $\left|\pm
      \frac{1}{2}\right> \leftrightarrow \left|\pm \frac{3}{2}\right>$
    and $\left|\pm \frac{3}{2}\right> \leftrightarrow \left|\pm
      \frac{5}{2}\right>$ hyperfine transition CW RHS spectra.  The
    solid lines represent the fit results and the shaded background
    the absolute value of the experimental spectra.  For each field
    orientation, the spectrum was normalized to the maximum signal
    amplitude.  }
\end{figure*}

Figure \ref{fig_groundstate} shows the experimental data for
$N_{tot}=101$ different external magnetic fields.
Several fit trails reliably led to the parameters shown in Table
\ref{tab_groundstate} and represented by the solid lines in Fig.\
\ref{fig_groundstate}.
\begin{table}
  \caption{Ground state spin Hamiltonian parameters \label{tab_groundstate}}
  \begin{ruledtabular}
    \begin{tabular}{crrc}
      parameter		&	value	&	fit error	&	unit	\\ \hline
      D			&	-6.3114	&	0.0027	&	MHz \\
      E			&	-0.8915	&	0.0021	&	MHz \\
      $\alpha_Q$		&	20.4		&	3.3		&	deg.	\\
      $\beta_Q$		&	147.7	&	1.4		&	deg.	\\
      $\gamma_Q$	&	10.2  	&	1.4		&	deg.	\\
      $g_x$			&	-51.7	&	3.6		&	MHz/T \\
      $g_y$			&	-23.5	&	1.1	  	&	MHz/T \\	
      $g_z$			&	-146.97	&	0.75		&	MHz/T \\
      $\alpha_M$	&	30.1		&	3.8		&	deg.	\\
      $\beta_M$		&	146.59	&	0.55		&	deg.	\\
      $\gamma_M$	&	13.09	&	0.69		&	deg.	\\
      $\alpha_{C_2}$		&	88.34	&	0.47		&	deg.	\\
      $\beta_{C_2}$		&	92.45	&	0.31		&	deg.	\\ \hline
    \end{tabular}
  \end{ruledtabular}
\end{table}

With these parameters, the RMS deviation between all accounted line
positions and the fit is $\approx 32$ kHz, significantly smaller than
the average linewidth of the ground state RHS lines of $\approx 196$
kHz (see Fig.\ \ref{fig_egspectra}), indicating that it is dominated
by statistical error. 
At the end of the fitting procedure, $L=$1218
of the total 1221 experimental lines from $N_{tot}=101$ spectra could
be assigned to calculated resonance line positions.  To estimate the
uncertainty of the fitted parameters, we sampled the parameter space
in the vicinity of the global minimum by repeating the probabilistic
part of the fitting procedure again using a fixed, low temperature.
Such a procedure can be shown to be rigorous if the only source of
error is gaussian noise in the line positions.
\cite{Longdell:2002p29,JosephLongdell:2003p1216} As an estimation for the
noise in the line positions we used the mean ratio of fitted line
widths $\sigma_i$ to the individual signal to noise (\it SNR\rm$_i$)
for all contributing RHS-lines:
\begin{equation}
  \nonumber
  \nu_\sigma = \frac{1}{L}\sum^{L}_i \frac{\sigma_i}{\mbox{\it SNR\rm}_i}.
\end{equation}
For the ground state data we found $\nu_\sigma=$ 1.4 kHz.  According to
this we chose the fixed temperature $T_{\sigma}$, so that a single
parameter change from its optimum value by $T_{\sigma_{l}}$ resulted in an
increase of the RMS deviation by $\nu_\sigma$.  After $2\cdot10^6$
iterations the histograms of the accepted parameters all showed a
gaussian shape, whose $1\sigma$-width are given as fit error in Table
\ref{tab_groundstate}.

Apart from the statistical error, we also consider systematic errors.
The most important contribution is due to the calibration error of the
magnetic field.  We estimate its precision to $\approx 0.65\%$, which
translates to the same fractional uncertainty of the gyromagnetic
ratios $g_x$, $g_y$ and $g_z$.  As the parameters are given in the
laboratory-fixed reference frame $(x,y,z)$ a misalignment of the
crystal does not contribute to the error but is expressed by the
$\alpha_{C_2}$ and $\beta_{C_2}$ values.  They determine the orientation of
the $C_2$ axis and therefore also that of the optical indicatrix
$(X,Y,Z)$ in our laboratory frame.  The only systematic contribution
in the angles arise from non-orthogonality of the coils, which is $<
1^\circ$.  The uncertainty of the alignment of the crystal relative to
our reference $(x,y,z)$ and that of the crystal surfaces to the
optical indicatrix $(X,Y,Z)$ results in an error of $\approx 5^\circ$
for the angles seen relative to the crystal axis system. 
The frequency scan was generated by direct digital synthesis,
resulting in negligible uncertainty in the frequency scale.

\subsection{\label{e}Excited state}

\begin{figure*}
  \includegraphics[width=2
  \columnwidth]{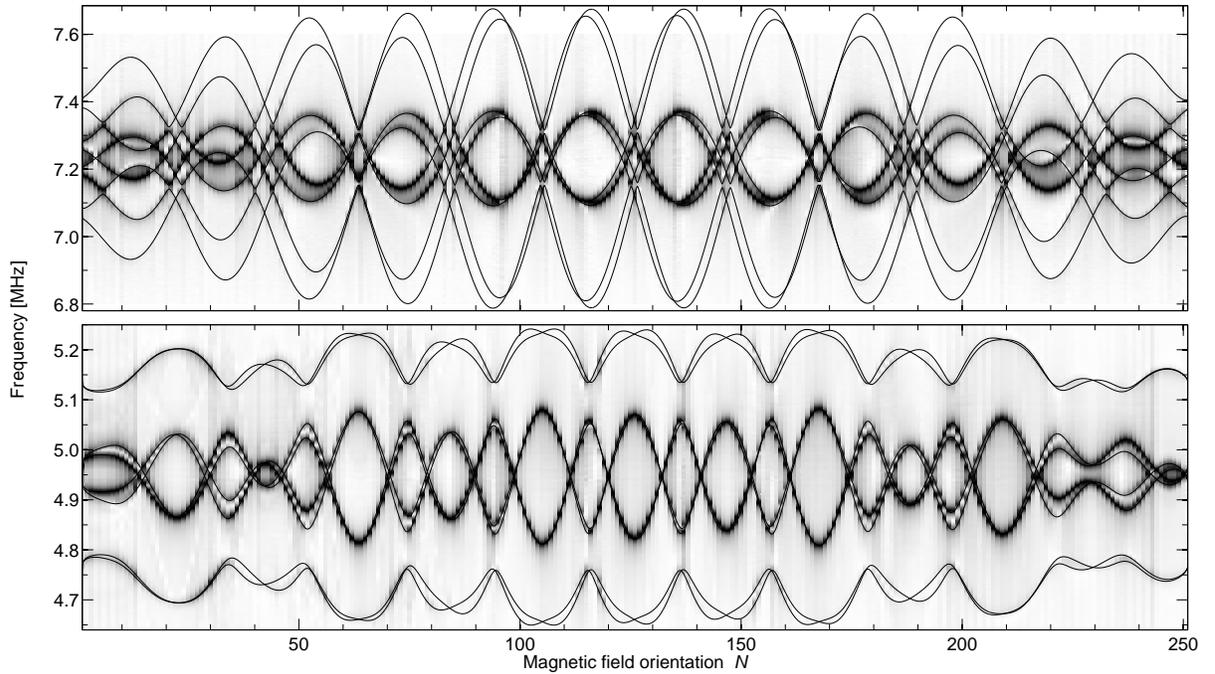}
  \caption{\label{fig_excitedstate} Excited state $\left|\pm
      \frac{1}{2}\right> \leftrightarrow \left| \pm
      \frac{3}{2}\right>$ and $\left|\pm \frac{3}{2}\right>
    \leftrightarrow \left|\pm \frac{5}{2}\right>$ hyperfine transition
    pulsed RHS spectra.  The solid lines represent the fit results and
    the shaded background the absolute value of the experimental
    spectra.  For each field orientation, the spectrum was normalized
    to the maximum signal amplitude.  }
\end{figure*}

For the excited state we used the same procedure as for the ground
state.  With the optimal fit parameters, we found an RMS deviation of
3.1 kHz between theoretical and experimental frequency values, using
$L=$ 2345 of the 2353 measured lines in $N_{tot}=251$ spectra.
Compared to the mean experimental FWHM of 22.3 kHz, the RMS deviation
is even better than for the ground state.  We mainly attribute this to
the higher quality of pulsed RHS spectra, with fewer line shape
artifacts.  Figure \ref{fig_excitedstate} and Table
\ref{tab_excitedstate} show the results.

\begin{table}
  \caption{Excited state spin Hamiltonian parameters \label{tab_excitedstate}}
  \begin{ruledtabular}
    \begin{tabular}{crrc}
      parameter		&	value	&	fit error	&	unit	\\ \hline
      D			&	1.90705	&	0.00023	&	MHz \\
      E			&	0.35665	&	0.00014	&	MHz \\
      $\alpha_Q$		&	-18.51	&	0.71		&	deg.	\\
      $\beta_Q$		&	73.83	&	0.48		&	deg.	\\
      $\gamma_Q$	&	-84.22	&	0.37		&	deg.	\\
      $g_x$			&	-17.22	&	0.27		&	MHz/T \\
      $g_y$			&	-14.39	&	0.10		&	MHz/T \\	
      $g_z$			&	-18.37	&	0.14		&	MHz/T \\
      $\alpha_M$	&	-23.7	&	2.4		&	deg.	\\
      $\beta_M$		&	88.5		&	4.1     	&	deg.	\\
      $\gamma_M$	&	-80.1	&	1.8		&	deg.	\\
      $\alpha_{C_2}$	&	88.63	&	0.25		&	deg.	\\
      $\beta_{C_2}$	&	92.69	&	0.24   	&	deg.	\\ \hline
    \end{tabular}
  \end{ruledtabular}
\end{table}
For the determination of the fit errors, we used the same procedure as
for the ground state.  The average uncertainty of the line positions
was $\nu_\sigma=$ 136 Hz.  The systematic errors are again dominated
by the calibration error of the magnetic field.  Although the crystal
was remounted between the excited and ground state experiments, the
resulting orientation of the $C_2$ axis agrees between the two data
sets within less than one degree, which is less than our alignment
accuracy.  As the Zeeman tensor is almost axially symmetric, $g_x
\approx g_z$, its orientation relative to the quadrupole
tensor or to the ground state tensor orientations can not be completely determined.

\subsection{\label{discussion}Discussion}
Examining Tables \ref{tab_groundstate} and \ref{tab_excitedstate}, it
appears that the principal values for the $\mathbf{Q}$ and
$\mathbf{M}$ tensors are similar to those found in other hosts like
\YSO\cite{Longdell:2002p29}, LaF$_3$ or YAlO$_3$
\cite{AAKaplyanskii:1987p1}. Especially, the ground state gyromagnetic
tensor is anisotropic with one large component, in contrast to the
excited state which also exhibits smaller values. To get some insight
into these properties, we compared these results with calculations
derived from crystal field calculations.

In a previous work \cite{GuillotNoel:2010p1250}, we found that
quadrupolar $D$ and $E$ values for ground and excited states (see Eq.\
(\ref{eq_Q})) could be very well reproduced starting from electronic
wavefunctions obtained by a crystal field analysis. The latter was
done assuming a $C_{2v}$ site symmetry, which is higher than the
actual one ($C_1$). In this higher symmetry, the $\mathbf{Q}$ tensors
for different crystal field levels are colinear, in clear
contradiction with our results. Nevertheless, it seems that the
additional crystal field parameters of $C_1$ symmetry have little
effects on the $\mathbf{Q}$ principal values. In the following, we
present the calculations of the $\mathbf{M}$ tensor principal values.

In $C_{2v}$ orthorhombic symmetry, the spin Hamiltonian of Eq.\
(\ref{eq_H}), expressed in the $(x_c,y_c,z_c)$ crystal field axes,
reads:
\begin{eqnarray}
  \label{eq:2}
  \mathcal{H''} = \sum_{i=x_c,y_c,z_c}&& B_ig_iI_i  \nonumber \\
  &&+ D({I^2_{z_c}}-\frac{I\left(I+1\right)}{3})
  +E({I^2_{x_c}}-{I^2_{y_c}})
\end{eqnarray}
where the $g_i$ are related to the $\mathbf{\Lambda}$ tensor by (see
Eq.\ (\ref{eq_H1})):
\begin{equation}
  \label{eq:3}
  g_i = -2A_Jg_J\mu_B\Lambda_{ii}-g_I\mu_N.
\end{equation}
The $\mathbf{\Lambda}$ tensor is given by Eq.\ (\ref{eq_lambda}) and
can be calculated from the electronic wavefunctions. The latter were
found using a free ion and crystal field Hamiltonian whose parameters
were fitted to experimentally determined crystal field levels
\cite{GuillotNoel:2010p1250}. In this calculation, we use arbitrary
permutations of the $(x_c,y_c,z_c)$ axes.  This results in different
sets of $E$,$D$ values which give the same hyperfine energy levels.
We subsequently fix the choice of the axis system such that the
convention $0 \le 3E/D=\eta \le 1$ [\onlinecite{Abragam:1994p1295}] is
fulfilled, thereby resolving ambiguous sets of parameters, such as (in MHz)
$D=2.082, E=-3.3445$ and $D=-6.0578, E=-0.6313$,
which describe identical ground state zero field hyperfine structures
and correspond to an exchange of $x_c$ with $z_c$.
With this convention, we fix the permutation of the axes and thus the values for  $D$ and $E$
for both electronic states.
The crystal field parameters we used and the corresponding $E$ and $D$ values
 for the ground and excited states are listed in Table \ref{tab:cf}.
\footnote{These parameters differ from those of Ref.\ \onlinecite{GuillotNoel:2010p1250} 
  which correspond to another assignement of crystal field axes.}  
The electronic wavefunction of the excited $^1$D$_2( \left|1  \right>)$ level 
is used instead of that of  $^1$D$_2( \left|0  \right>)$ to take into account 
a wrong ordering in the calculated crystal field levels of this
multiplet \cite{Esterowitz:1979p1249,GuillotNoel:2010p1250}.

\begin{table}
  \centering
  \caption{Crystal field parameters $B_{ij}$,  calculated
    $\mathbf{Q}$ and  $\mathbf{M}$ tensor principal values
    and second order hyperfine interaction parameters.} 
  \begin{ruledtabular}
    \begin{tabular}{cr|crrc}
      $B_{ij}$ 	&	(cm$^{-1}$)			&					&	ground state	&	excited state	&	unit  		\\ \hline		 
      $B_{20}$	&    	375 					&   	$D$				& 	-6.1			& 	2.0			&	MHz	 	\\
      $B_{22}$	&   	-93  					&	$E$  			&  	-0.63		&      0.30         		&	MHz 	\\ 
      $B_{40}$	&	768					&     	$g_{x_c}$			&  	-32			&      -18        		&	MHz/T	\\
      $B_{42}$	&	445 					&    	$g_{y_c}$			&     	-22			&      -4         		&	MHz/T	 \\
      $B_{44}$	&	1027					&    	$g_{z_c}$  		&  	-151  		& 	-18			&     	MHz/T	\\
      $B_{60}$	&	267 					&    	$A_J$ 			&  	937			&	697			&	MHz		 \\
      $B_{62}$	&	-402					&  	$g_J$ 			&	0.81			&	1.03			&			 \\
      $B_{64}$	&	-61					&   $\Lambda_{x_cx_c}$	&	0.0280		& 	0.0066		& 	cm		\\
      $B_{66}$	&	-52    				&   $\Lambda_{y_cy_c}$	& 	0.0140		&	-0.0090		&	cm		\\
      			&   						&   $\Lambda_{z_cz_c}$ 	& 	0.2000		& 	0.0069		&	cm		\\
    \end{tabular}
  \end{ruledtabular}
  \label{tab:cf}
\end{table}

Comparing Tables \ref{tab_groundstate} and \ref{tab_excitedstate} with
Table \ref{tab:cf} shows that a reasonable agreement is found between
experimental and calculated principal values of ground and excited
state $\mathbf{M}$ tensors. 
Again this suggests that additional parameters appearing in
calculations using $C_1$ symmetry mainly determine the relative
orientation between the different tensors.  Calculated values
especially reproduce two features mentioned above: the very large
value of $g_z$ for the ground state and the smaller $g_i$ values for
the excited state compared to the ground state.  A qualitative
understanding of these properties can be obtained from the crystal
field analysis by looking at the different factors entering in Eq.\
(\ref{eq:3}). They are summarized in Table \ref{tab:cf}. We first note
that the isotropic and crystal field independent nuclear Zeeman
contribution to $g_i$ equals -12.2 MHz/T. Differences in $g_i$
values are mainly linked to the pseudoquadrupole $\mathbf{\Lambda}$
tensors since the products $A_Jg_J$ vary by only 4\% between ground
and excited states. The pseudoquadrupole
tensors involve the $\vec{J}$ matrix elements
and the energy differences appearing as denominators in Eq.\
(\ref{eq_lambda}). We first discuss the ground state case. The
electronic wavefunction of interest (the lowest energy crystal field
level) has the following form:
\begin{equation}
  \nonumber
  \left|0  \right> = -0.62  \left|^3H_4,-4  \right> 
  -0.62\left|^3H_4,4  \right> -0.4 \left|^3H_4,0  \right>
\end{equation}
where brackets on the right hand side are written as
$\left|^{2S+1}L_J,M_J \right>$ and only terms with a coefficient
larger than 0.15 have been kept.  The larger $J_{z_c}$ matrix
element is found between $\left|0 \right>$ and $\left|1 \right>$,
since the latter is nearly only composed of $\left|^3H_4,\pm 4 \right>
$ states. The $ \left|\left<0 \left|J_{z_c} \right|1 \right>\right|$
matrix element equals 3.6 close to the maximum value of $\left|
  \left<^3H_4,\pm 4 \right|J_{z_c} \left|^3H_4,\pm 4
  \right>\right|=4$.  Moreover, this large matrix element is found for
levels close in energy (65 cm$^{-1}$ Ref.\
\onlinecite{GuillotNoel:2010p1250}), resulting in a large $\Lambda_{z_cz_c}$.  On
the other hand, $\left|0\right>$ couples to crystal field levels containing
$\left|^3H_4,\pm 3 \right>$, $\left|^3H_4,\pm 1 \right>$ by $J_{x_c}$
or $J_{y_c}$ operators.  The corresponding matrix elements do not
exceed 2.4 in absolute value.  As expected,
this is close to the average value of matrix elements of the form
$\left<^3H_4,\pm 4 \right|J_{i} \left|^3H_4,\pm 3 \right>$ and
$\left<^3H_4,\pm 1 \right|J_{i} \left|^3H_4,\pm 2 \right>$ (where
$i=x_c$ or $y_c$), which is at most 1.9. The
levels with the largest matrix elements are located at high energies
($E_3=143$ and $E_5=349$ cm$^{-1}$ for $J_{y_c}$ and $J_{x_c}$
respectively), resulting in low $\Lambda_{x_cx_c}$ and
$\Lambda_{y_cy_c}$. This in turn explains the small values of
$g_{x_c}$ and $g_{y_c}$ compared to $g_{z_c}$.

A similar analysis can be performed for the \odt{} excited state. 
The level of interest is $\left|1\right>$ because the crystal field
calculation inverts levels $\left|0\right>$ and $\left|1\right>$ as mentioned above.
The latter is found to be equal to:
\begin{equation}
  \nonumber
  \left|1\right> = 0.67
  \left|
    ^1D_2,-2
  \right>-0.67
  \left|
    ^1D_2,2
  \right>
\end{equation}
with the same convention as above. This state gives a $J_{z_c}$ matrix
element equals to 2 with the state $\left|4\right>$, located 441
cm$^{-1}$ higher than $\left|1\right>$. Maximum average values for
matrix elements of $J_{x_c}$ and $J_{y_c}$ can be estimated as above
for levels containing $\left|^1D_2 \pm1 \right>$ states, resulting in
$\left|\left<0 \left|J_{x_c} \right|1\right>\right| \approx
\left|\left<1 \left|J_{x_c} \right|2\right>\right| \approx 1$.  The
corresponding energies are $E_0-E_1=-82$ cm$^{-1}$ and \mbox{$E_2-E_1=113$
cm$^{-1}$.} The combination of matrix elements and energy differences
result in smaller values for $\Lambda_{ii}$ ($i=x_c,y_c,z_c$) compared
to the ground state. This can also partly explain the isotropy of the
excited state $g_i$ values, which are closer to the nuclear Zeeman
contribution. As pointed out above, several \pr doped compounds
exhibit the same behavior so that the discussion given above could
also be applied to them.

We now turn to the principal axes of the spin Hamiltonian tensors.  As
a further test of the ($\alpha_Q,\beta_Q,\gamma_Q$) parameters
determined from the RHS experiments, we compared the relative
oscillator strengths obtained from zero field spectral tayloring
experiments \cite{GuillotNoel:2009p1163} with calculations. The
oscillator strengths are assumed to be proportional to the square of
the overlap of the nuclear wavefunctions \cite{AAKaplyanskii:1987p1},
the latter being given by the ground and excited state Hamiltonians.
This assumption is reasonable since the hyperfine interactions are a
small perturbation to the electronic wavefunctions. The results are
gathered in Table \ref{tab:osc}. A good agreement is found,
showing that indeed the orientation of the quadrupole tensors was
determined correctly. In \pr{}:\YSO, significant discrepancies were found
between calculated and experimental values \cite{Nilsson:2004p25, Nilsson:2005p1315}. 
This was tentatively attributed to additional selection rules due to
superhyperfine coupling with Y ions.  In our case, it seems that
although superhyperfine coupling may also be observed (see Sec.\
\ref{ZEFOZ}), relative optical transition matrix elements can still be
determined from the overlap of the nuclear wavefunctions.

Hyperfine transition linewidths were also determined during the fit procedure (Fig.\ 3).
The data show that the transitions with the larger splittings also show the larger linewidths. 
For example, the ground state \mbox{$\left | \pm \frac{3}{2}\right> \leftrightarrow \left |\pm \frac{5}{2} \right>$} 
transitions at 24.44 MHz have an average linewidth of \mbox{301 kHz} (see Fig.\ 3, caption) 
whereas the \mbox{$ \left| \pm \frac{1}{2}\right> \leftrightarrow \left |\pm\frac{3}{2}\right>$} transitions at 
14.87 MHz have a linewidth of only \mbox{105 kHz.} To explain this, we first consider 
that the used fields are small enough, so that the observed linewidths are similar 
to those obtained at zero field. Moreover, we approximate the spin Hamiltonian by 
setting $E = 0$ in Eq.\ (\ref{eq:2}) so that $\mathcal{H''} = D(I_{z_c}^2-I(I+1)/3)$. 
In this case, the $ \left| \frac{1}{2}\right> \leftrightarrow \left |\frac{3}{2}\right>$ transition energy is $\left|2D\right|$
and the $ \left| \frac{3}{2}\right> \leftrightarrow \left |\frac{5}{2}\right>$ is $\left|4D\right|$.
Crystal field variations from one ion position to an other correspond to a distribution of crystal field 
parameters and therefore of the $D$ parameter. The hyperfine linewidths in the excited 
and ground state should then be proportional to the transition energies. 
This is qualitatively in agreement with the experimental values. 
The excited state linewidths are also smaller than the ground state ones, 
which suggests that the $D$ distribution width is also proportional to $D$.

\begin{table}
  \centering
  \caption{Experimental and calculated relative optical oscillator strengths
    between \thf{} and \odt{}  hyperfine levels. 
    Rows correspond to transitions starting from the ground state 
    hyperfine levels and columns correspond to transitions to different
    excited state hyperfine levels (see Fig.\ \ref{fig_levels}).
    Experimental data from Ref.\ \onlinecite{GuillotNoel:2009p1163}.}

  \begin{ruledtabular}
    \begin{tabular}{ccccc}
      & & & \\
      &     		&\raisebox{1.5ex}[-1.5ex]{$\left|e,\pm\frac{1}{2}\right>$}
      &\raisebox{1.5ex}[-1.5ex]{$\left|e,\pm\frac{3}{2}\right> $}& \raisebox{1.5ex}[-1.5ex]{$\left|e,\pm\frac{5}{2}\right> $} \\ \hline{}
      & exp 	& $0.09\pm0.01$ 	& $0.28\pm0.01$ 	& $0.63\pm0.01$ \\
      \raisebox{1.5ex}[-1.5ex]{$\left<g,\pm\frac{1}{2}\right| $}	& cal 	& $0.08\pm0.01$        		& $0.24\pm0.02$        		& $0.67\pm0.02$        \\ \\
      & exp 	& $0.33\pm0.01$ 	& $0.39\pm0.01$ 	& $0.28\pm0.02$ \\
      \raisebox{1.5ex}[-1.5ex]{$\left<g,\pm\frac{3}{2}\right| $} 	& cal 	& $0.31\pm0.02$        		& $0.45\pm0.02$        		& $0.24\pm0.02$        \\ \\
      & exp 	& $0.55\pm0.01$ 	& $0.36\pm0.01$ 	& $0.09\pm0.01$ \\
      \raisebox{1.5ex}[-1.5ex]{$\left<g,\pm\frac{5}{2}\right| $}	& cal 	& $0.60\pm0.02$        		& $0.31\pm0.02$        		& $0.09\pm0.01$        \\   
    \end{tabular}
  \end{ruledtabular}
  \label{tab:osc}
\end{table}

\subsection{\label{ZEFOZ} Experimental verification of a ZEFOZ transition}
\begin{figure}
  \includegraphics[width=\columnwidth]{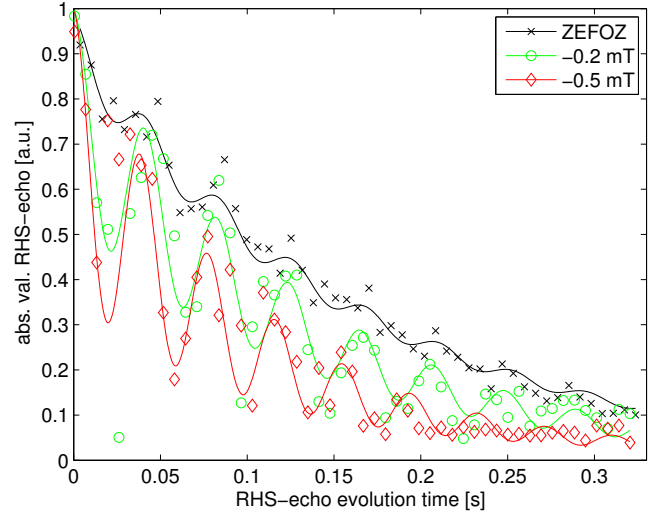}
  \caption{\label{fig_ZEFOZ} (Color online) Raman-echo decays at ZEFOZ point and with
    magnetic detunings of -0.2 and -0.5 mT for the z-component.}
\end{figure}

As mentioned earlier the coherence times for ZEFOZ transitions are
expected to be much longer than at zero or arbitrary magnetic field.
Up to date this was demonstrated experimentally only for
\pr:\YSO{}.\cite{Fraval:2004p115,Fraval:2005p122} To proof the
usefulness of the ZEFOZ technique for other compounds and also to
verify our hyperfine characterization we present here experimental
data of a ZEFOZ transition of \pr:\LAWO{} in the following.  Using our
parametrization of the spin Hamiltonian we sought for magnetic field
configurations and transitions that satisfy the ZEFOZ conditions
\cite{Fraval:2004p115}:
\begin{equation}
  \label{eq_FOZ}
  \vec{S}^I(\vec{B_{opt}}) = \left(
    \begin{array}{ccc}
      \frac{\partial\nu_i(\vec{B}_{opt})}{\partial B_x},	 &
      \frac{\partial\nu_i(\vec{B}_{opt})}{\partial B_y},	 &
      \frac{\partial\nu_i(\vec{B}_{opt})}{\partial B_z}			
    \end{array}\right) = \vec{0}.
\end{equation}
We identified such points by numerical minimization of
$|\vec{S}^I(\vec{B})|$ for all transitions $\nu_i$ within a static magnetic
field grid.  In this way we found several ZEFOZ positions where one
transition satisfies Eq.\ (\ref{eq_FOZ}) and further showing low
curvature, e.g. small second order coefficients
\begin{equation}
  \label{eq_SII}
  S^{II}_{jk}(\vec{B}) = \left. \frac{\partial^2 \nu_i(\vec{B})}{\partial B_j \partial B_k} \right|_{\vec{B}}.
\end{equation}
An identified (hyperfine ground state) ZEFOZ transition at
$\nu_4=12.6$ MHz and $\vec{B}_{opt}=($57.5, 4.0, -36.1$)$ mT was
experimentally explored.  The setup at TU Dortmund described in
Section \ref{experiment} was not designed for magnetic fields of more
than 12 mT per axis.  Experiments exploring the ZEFOZ point we carried
out in Lund. This offered the opportunity to experimentally verify the
Hamiltonian parameters and predicted ZEFOZ points in an independent
laboratory.  The static magnetic field vector was provided by a set of
three orthogonal superconducting coils, the y-coil being part of an
Oxford Spectromag
cryogenic 7 T magnet assembly with 0.1 mT resolution.  The
homebuilt $x$ and $z$ coils could generate fields of a few \mbox{100 mT} and
were controlled by 16 bit DAC.  To fit into the homogeneous region of
the coils we had to cut a 5x5x1 mm piece from the sample that was used
for the characterization.  Due to this and the construction of the
sample holder, we could only align the optical indicatrix with high
precision along the $z$-axis (laser direction) of the coil frame. The
other axis orientations were only known with a precision of about 10
degrees.  To find the ZEFOZ point experimentally we had to consider
this misalignment, the accuracy of the Hamiltonian parameters and the
calibration of the coils.  Therefore in a first step we adjusted the
magnetic field to get a good overlap between observed CW RHS spectra
and the calculated line positions, that follow from Table
\ref{tab_groundstate} and the ZEFOZ field.  As the transition of
interest shows very small frequency changes when being close to the
desired field vector, we utilized a $T_2$ measurement by Raman-echos,
induced by two RF pulses ($P\approx 3$ W, $\tau_p \approx 25/50$
$\mu$s, pulses along z-axis), in a second step.  Thus we fine tuned
the magnetic field components for maximum Raman-echo signal at long
echo times (RF pulse separations).  Figure \ref{fig_ZEFOZ} shows the
longest-lived Raman-echo decay curves we could achieve.  These
demonstrate hyperfine coherence times $T_2(\vec{B}_{opt})$ of up to
$158\pm7$ ms, representing a 630 fold increase compared to the zero
magnetic field situation \cite{Goldner:2009p726}.
The decay curves at magnetic fields slightly detuned 
from the ZEFOZ point show slow modulations with a frequency of
$\nu_M=24.5\pm1.7$ Hz. This could
be due to a superhyperfine interaction with La nuclei, 
but a clear explanation is lacking at the present time.
This point will be investigated in further experiments.
To fit decay times we used the function
\begin{equation}
  \nonumber
  f(t) = \left(A+M\cos\left(2\pi\nu_Mt\right) \right)\exp\left({\frac{-t}{T_2}}\right)+c,
\end{equation}
where $A$ is the exponentially decaying part, $\nu_M$ is the
modulation frequency, $T_2$ the decay time and $c$ an offset.  As we
moved the magnetic field away from the ZEFOZ point $\vec{B}_{opt}$ by $-0.2$
mT in the z-component this resulted in a decrease of the coherence
time to $T_2(-0.2\mbox{ mT})=133\pm16$ ms and a shift by $-0.5$ mT
resulted in $T_2(-0.5\mbox{ mT})=97\pm19$ ms. 

In the following we will use this measured $T_2(\vec{B})$
values to estimate the magnetic field fluctuation $\Delta B$ at the \pr{} site.
These fluctuations cause frequency shifts and thereby broaden the hyperfine transition.
For this purpose, we expand the hyperfine transition frequency $\nu_i$ on a deviation $B_{off}$
from a given field $\vec{B}$ 
\begin{equation}
\nu_i\left|_{\vec{B}} \right. = \nu_i(\vec{B}) + s1 \, B_{off} + \frac{s2}{2} \, B_{off}^2. \nonumber
\end{equation}
If the reference field $\vec{B}$ fulfills the  ZEFOZ condition, the first derivative vanishes, $s_1=0$ and 
the frequency shift due to the fluctuations is
\begin{equation}
\Delta \nu =  \frac {s2}{2} (\Delta B)^2. \label{dnu}
\end{equation}
Since the fluctuations only generate positive frequency shifts (for $s2 > 0$), they add a line broadening of $\Delta \nu / 2$.
Using the assumption that the line broadening is entirely due to magnetic field fluctuations\cite{Longdell:2006p9},
we obtain a decay rate
$$
T_2^{-1} =  \frac {s2}{4} (\Delta B)^2.
$$
With the experimentally obtained value of $T_2=158$ ms and $s2 \approx12$ kHz/mT$^2$, 
calculated from the maximum eigenvalue of the derivative matrix Eq.\ (\ref{eq_SII}),
using the parameters from Table \ref{tab_groundstate},
we thus estimate  the magnetic field fluctuations as $\Delta B \approx 46$ $\mu$T.
This value is of the same order of magnitude as that found for the ZEFOZ points in \pr:\YSO{}, the only ones 
experimentally investigated up to date, where  $s2 \approx3$-$6$ kHz/mT$^2$ and $\Delta B=14$ $\mu$T
\cite{Fraval:2005p122,Longdell:2006p9} were found.
The zero field relaxation times in \pr:\LAWO{} and  \pr:\YSO{} are also comparable ($\approx250$ $\mu$s vs. $\approx500$ $\mu$s).

We now analyze the dependence of the relaxation times on the magnetic field when the deviation is large compared to
the amplitude of the fluctuations, $B_{off} \gg \Delta B$.
This changes Eq. (\ref{dnu}) to
\begin{equation}
\Delta \nu =   \frac{s2}{2} (B_{off}^2 + 2 \Delta B \, B_{off}). \nonumber
\end{equation}
The first term describes the line shift, the second a line broadening. 
Using the value  $\Delta B\approx 46$ $\mu$T for the magnetic field fluctuations,
we expect that the resulting line broadening for a field change of $B_{off}= 0.2$ mT
is \mbox{$\approx 9$ ms} and for $B_{off}= 0.5$ mT $\approx 3.6$ ms.
These values are significantly shorter than the experimental values. 

The most likely explanation for this discrepancy is that the relaxation at our reference field
is not entirely due to magnetic field fluctuations and that the reference field does not exactly
fulfill the ZEFOZ condition. 
Both effects lead to additional contributions to the dephasing rate. We therefore write the total
dephasing rate as
\begin{equation}
T_2^{-1} = T_{2,0}^{-1} + s1 \, \Delta B + s2 \, \Delta B \, B_{off}. \label{T2Dep}
\end{equation}
Here, $T_{2,0}^{-1}$ describes those contributions that are not due to magnetic field fluctuations, such as
phonons, while $s1$ is the first derivative of the transition frequency with respect to the magnetic field change.
Both contributions to the dephasing rate are independent of the magnetic field offset $B_{off}$ and therefore
not distinguishable in the available experimental data.

Using now $s2 \approx 8.2$ kHz/mT$^2$, corresponding to the projection of Eq.\ (\ref{eq_SII}) into the 
the $z$-direction, we use Eq.\ (\ref{T2Dep}) to estimate the magnetic field fluctuations.
The result of a linear fit of $T_2^{-1}$ vs. $B_{off}$ yields \mbox{$\Delta B \approx 1$ $\mu$T}. 
Applying the same analysis to the  \pr:\YSO{} data (estimated from 
Figure 2 in Ref.\ \onlinecite{Fraval:2005p122}) gives \mbox{$\Delta B \approx 1.3$ $\mu$T}.
This also reduces the estimate for $\Delta B$ by approximately one order of magnitude compared to the analysis
where the minimum dephasing rate is assumed to originate entirely from the quadratic term of the magnetic
field fluctuations \cite{Longdell:2006p9}.

To get a better estimate for $\Delta B$ further measurements would be required.
The phononic contributions to $T_2$ could be determined from measurements at different temperatures, 
as $s1$ and $s2$ are independent of this parameter. 
Quantitative measurements of $T_2$ at several deviations $B_{off}$ in three directions could help 
to estimate the numerical value of the second term of Eq.\ (\ref{T2Dep}).

\section{\label{conclusion}Conclusion}
We characterized the hyperfine interaction of praseodymium doped into \LAWO{}  for the electronic
ground state and one electronically excited state. 
We described in detail the experimental and numerical methods to 
reliably derive the spin Hamiltonian parameters.
The relative oscillator strengths between the \thf{} and \odt{} hyperfine levels derived from our data are in 
good agreement with those measured in earlier experimental work \cite{GuillotNoel:2009p1163}.
We indicated physical reasons for the experimentally found tensor orientations 
and their principal axis values by a crystal field analysis discussion. Further the calculated tensor
values following from this analysis are in reasonable agreement with the values derived from the
experimental data, backing up our results and showing the usefulness of such analysis techniques.
The full characterization enabled us to calculate all transition frequencies for arbitrary magnetic fields. 
Using this, we could predict the magnetic field value at which a ZEFOZ transition occurs.
We verified this condition experimentally in a second laboratory.
Besides the experimental verification of the hyperfine characterization, we determined
the order of magnitude of the magnetic fluctuations at the \pr-site ($\Delta B \lesssim 10$ $\mu$T) and investigated
the coherence properties of the $\mbox{ZEFOZ}$ transition.
The latter showed characteristics similar to superhyperfine interaction and 
a up to 630-fold increase of the coherence lifetime compared to zero field. 
The demonstrated spin lifetime of 158 $\pm$ 7 ms and the relatively low second order 
Zeeman coefficient show that even crystal systems with high magnetic moment density 
can have a high potential for quantum memory and information applications.

\section*{Acknowledgments}
The authors are grateful to J. J. Longdell and M. J. Sellars for useful discussions 
in preparation of the ZEFOZ measurements. 
This work was supported by the Swedish Research Council, 
the Knut \& Alice Wallenberg Foundation, the Crafoord Foundation 
and the EC FP7 Contract No. 228334 and 247743 (QuRep).

\appendix*
\section{Euler angles conventions \label{App_EulerAngs}}
For the tensor transformations we used the common ``zyz''-convention
\cite{Goldstein:2002p1241} with right-handed coordinate systems.  The
transformation matrices $R$, e.g. in Eq.\ \ref{eq_M} and \ref{eq_Q},
are given by:
\begin{widetext}
  \begin{equation}
    A = \left( 
      \begin{array}{ccc}
        \cos \alpha		& \sin \alpha 	& 		0 \\
        -\sin \alpha		& \cos \alpha	& 		0 \\
        0				&			& 		1
      \end{array} 
    \right), B =  \left(
      \begin{array}{ccc}
        \cos \beta		& 0		 	&-\sin \beta \\
        0				& 1			& 		0 \\
        \sin \beta		& 0			& \cos \beta
      \end{array} 
    \right), 
    C =	 \left(
      \begin{array}{ccc}
        \cos \gamma	& \sin \gamma 	& 		0 \\
        -\sin \gamma	& \cos \gamma	& 		0 \\
        0				&			& 		1
      \end{array} 
    \right) \nonumber 
  \end{equation}
  \begin{equation}
    R\left( \alpha, \beta, \gamma \right) = C \cdot B \cdot A =  \left(
      \begin{array}{ccc}
        -\sin \alpha   \sin \gamma 	+ \cos \alpha   \cos \beta   \cos \gamma	& \cos \alpha  \sin \gamma + \cos \beta  \cos \gamma  \sin \alpha 	&  -\cos\gamma  \sin \beta \\
        -\cos \gamma  \sin \alpha 	- \cosÊ\alpha   \cos \beta  \sin \gamma	& \cos \alpha Ê\cosÊ\gamma- \cos \beta  \sin \alpha  \sin \gamma 	&  \sin \beta  \sin \gamma   \\
        \cos \alpha  \sin \beta										& \sin \alpha  \sin \beta									&  \cos \beta
      \end{array}
    \right),
    \label{eq_EulerDef}
  \end{equation}
\end{widetext}
with $\alpha$, $\gamma$ $\epsilon$ $[-\pi,\pi]$ and $\beta$ $\epsilon$ $[0,\pi]$.

\end{document}